\documentclass[aps,pra,twocolumn,showpacs]{revtex4-1}
\usepackage[matrix,frame,arrow]{xypic}
\usepackage[matrix,frame,arrow]{xy}
\usepackage{amsmath}
\newcommand{\bra}[1]{\left\langle{#1}\right\vert}
\newcommand{\ket}[1]{\left\vert{#1}\right\rangle}
\newcommand{\qw}[1][-1]{\ar @{-} [0,#1]}
\newcommand{\qwx}[1][-1]{\ar @{-} [#1,0]}

\newcommand{\measureD}[1]{*{\xy*+=+<.5em>{\vphantom{\rule{0em}{.1em}#1}}*\cir{r_l};p\save*!R{#1} \restore\save+UC;+UC-<.5em,0em>*!R{\hphantom{#1}}+L **\dir{-} \restore\save+DC;+DC-<.5em,0em>*!R{\hphantom{#1}}+L **\dir{-} \restore\POS+UC-<.5em,0em>*!R{\hphantom{#1}}+L;+DC-<.5em,0em>*!R{\hphantom{#1}}+L **\dir{-} \endxy} \qw}

\newcommand{\control}{*!<0em,.025em>-=-{\bullet}}
\newcommand{\controlo}{*-<.21em,.21em>{\xy *=<.59em>!<0em,-.02em>[o][F]{}\POS!C\endxy}}

\newcommand{\qswap}{*=<0em>{\times} \qw}
\newcommand{\multigate}[2]{*+<1em,.9em>{\hphantom{#2}} \qw \POS[0,0].[#1,0];p !C *{#2},p \save+LU;+RU **\dir{-}\restore\save+RU;+RD **\dir{-}\restore\save+RD;+LD **\dir{-}\restore\save+LD;+LU **\dir{-}\restore}
\newcommand{\ghost}[1]{*+<1em,.9em>{\hphantom{#1}} \qw}

\newcommand{\ustick}[1]{*!D!<0em,-.5em>=<0em>{#1}}

\newcommand{\Qcircuit}[1][0em]{\xymatrix @*[o] @*=<#1>}

\newcommand{\pureghost}[1]{*+<1em,.9em>{\hphantom{#1}}}

\newcommand{\poloFantasmaCn}[1]{{{}^{#1}_{\phantom{#1}}}}

\usepackage{epsfig}
\usepackage{amsbsy,latexsym}
\usepackage{amsmath}
\usepackage{amssymb, mathrsfs}
\usepackage[mathscr]{eucal}
\usepackage{hyperref}

\def\d{\mathrm{d}}
\def\<{\langle}
\def\>{\rangle}
\def\Tr{\operatorname{Tr}}

\def\:{\hbox{\bf :}}

\def\set#1{{\sf #1}}
\def\dag{\dagger}

\def\leq{\leqslant}
\def\map#1{\mathcal #1}
\def\sH{\mathcal{H}}

\def\supp{\set{Supp}}
\def\qed{$\,\blacksquare$\par}

\def\Cmplx{\mathbb C}
\newcommand{\Ket}[1]{| #1 \rangle \! \rangle}
\newcommand{\Bra}[1]{\langle \! \langle #1 |}
\newcommand{\BraKet}[2]{\langle \! \langle #1 | #2 \rangle  \! \rangle}
\newcommand{\KetBra}[2]{\Ket{#1} \Bra{#2}}
\newcommand{\braket}[2]{\langle #1 | #2 \rangle}
\newcommand{\ketbra}[2]{\ket{#1} \bra{#2}}
\newcommand{\hilb}[1]{\mathcal{#1}}

\newtheorem{theorem}{Theorem}
\def\Proof{\medskip\par\noindent{\bf Proof. }} 

\begin{document}
\title{Information-Disturbance Tradeoff in the Estimation of a Unitary Transformation}

\author{Alessandro Bisio}
\affiliation{QUIT group,  Dipartimento di Fisica ``A. Volta'', INFN Sezione di Pavia, via Bassi
  6, 27100 Pavia, Italy} 
\author{Giulio Chiribella} 
\affiliation{Perimeter Institute for Theoretical Physics, 31 Caroline St. North, Waterloo, Ontario N2L 2Y5, Canada } 
\author{Giacomo Mauro D'Ariano}
\author{Paolo Perinotti} 
\affiliation{QUIT group,  Dipartimento di Fisica ``A. Volta'', INFN Sezione di Pavia, via Bassi
  6, 27100 Pavia, Italy.}

\date{ \today}
\begin{abstract} 

We address the problem of the information-disturbance trade-off associated to the estimation of a quantum transformation, and show how the extraction of information about the a black box causes a perturbation of the corresponding input-output evolution.  
In the case of a black box performing a unitary transformation, randomly distributed according to the invariant measure, we give a complete solution of the problem, deriving the optimal trade-off curve and presenting an explicit construction of the optimal quantum network.      
\end{abstract}
\maketitle

\section{Introduction}
One of the key features of quantum theory is the impossibility of extracting information from a system without producing
disturbance on its state, the only exception to this rule being the trivial case when the state belongs to a set of orthogonal states. A canonical illustration of the unavoidable disturbance caused by  quantum measurements is Heisenberg's  $\gamma$-ray microscope thought experiment  \cite{heisenberg}.
The impossibility of a non-disturbing extraction of information is the working principle of quantum cryptography, whose security relies on the fact that any amount of information extracted by the eavesdropper causes a corresponding amount of disturbance that can be detected by the communicating parties.    
A quantitative expression of such information-disturbance trade-off is a non-trivial 
issue because there are many  different ways to quantify ``information" and ``disturbance", which have been put forward in the literature  \cite{scully,stenholm,englert,fuchs-peres, barnum, banaszek, d'ariano, ozawa, fiur, ent-tradeoff, dema, macca, werner, busc, facco}.  

All the scenarios analyzed in the past have one point in common: they concern the disturbance produced by measurements on \emph{quantum states}.  However, one can consider other scenarios where the measurements produce a disturbance on \emph{quantum transformations}.  For example, we may have a black box implementing an unknown transformation belonging to a set $\{\map E_i\}$, with the restriction that the black box can be used only one time.  
On the one hand, we may try to identify the unknown transformation (that is,  to find out the index $i$). On the other hand, we may want to use the black box on a variable input state.   Clearly, in general the two tasks are incompatible:  In this case there is a trade-off between the amount of information that can be extracted about a black box and the disturbance caused on its action. In other words, we cannot estimate an unknown quantum dynamics without perturbing it.    Therefore, it is important to find a quantitative formulation of the information-disturbance trade-off, and to find the optimal scheme that introduces the minimum amount of disturbance for any given amount of extracted information.   
Like the trade-off for states,  the trade-off for transformations is relevant to the discussion of quantum cryptographic protocols where the secret key is encoded in a set of transformations, as it happens in the  two-way protocols of Refs. \cite{pingpong,mancini,clonunit} for finite dimensional systems, and in the protocol of Ref.  \cite{pirandola} for continuous variables.  Here for simplicity we will restrict our attention to the case of unitary transformations on finite dimensional quantum systems.  Like in Refs. \cite{banaszek,ent-tradeoff} we will quantify the information gain and the disturbance with suitable fidelities, and we will derive the minimum amount of disturbance associated to any  possible value of the information gain.      

The paper is structured as follows:  in Section \ref{sec:prelim} we introduce the problem and the notation.   Section \ref{sec:combs} then provides a brief review of the formalism of quantum combs and generalized instruments \cite{architecture,supermaps,comblong}, which is crucial in our paper.  The complete analysis of the information-disturbance trade-off for arbitrary unitary transformations will be presented in Section \ref{sec:tradeoff}:  in particular,  we will first give the rigorous mathematical formulation of the problem (Subsec. \ref{subsec:formulation} ), the analysis of its symmetries \ref{subsec:symmetries}, the derivation of the optimal trade-off curve (\ref{subsec:curve}), and, finally, the construction of the optimal network (\ref{subsec:network}).  We conclude the paper with a discussion of the results in Section \ref{sec:conclusions}.          
 
\section{Preliminaries and notation}\label{sec:prelim}


In the case of states, the mathematical  tool to analyze the information-disturbance trade-off is provided by the notion of \emph{quantum instrument}.
In the discrete-outcome case, a quantum instrument is a set of \emph{quantum operations} (trace-decreasing completely positive maps) $\{ \mathcal{T}_i  \} $  transforming operators on the input system Hilbert space $\hilb H_0$  to operators on the output system Hilbert space $\hilb H_1$, with the normalization condition   that $\map T :=  \sum_i \map T_i$ is trace-preserving (that is, it is a \emph{quantum channel}).        A quantum instrument  describes a measurement process that outputs the classical outcome $i$  and  the  quantum state
$ \mathcal{T}_i (\rho) / \Tr[ \mathcal{T}_i (\rho)]$ with probability $p_i = \Tr[ \mathcal{T}_i (\rho)]$.  To derive our results we will use the generalization of the notion of instrument to measurement processes on quantum transformations, rather than on quantum states  \cite{architecture,supermaps,comblong}.  This extension will be presented in Section \ref{sec:combs}.

 In the following we will denote the linear operators on a Hilbert space $\hilb H$ by $L(\hilb H)$.  We will make extensive use of the isomorphim between linear operators in $L(\hilb{H})$
and vectors in $\hilb{H} \otimes \hilb{H}$ given by
\begin{equation} \label{doubleketeq}
A = \sum_{nm}\bra{n}A\ket{m} \ketbra{n}{m} 
\leftrightarrow
\Ket{A} = \sum_{nm}\bra{n}A\ket{m} \ket{n}\ket{m} 
\end{equation}
where $\{ \ket{n} \}$ is a fixed orthonormal basis for $\hilb H$.   The isomorphism satisfies the property 
\begin{equation}
\Ket A  =  (A \otimes I)  \Ket I  =  (I \otimes A^T)  \Ket I 
\end{equation}
$T$ denoting transposition with respect to the fixed basis. 
For the sake of clarity we will often use the notation  $\hilb{H}_{a,b} $ to denote the tensor product $ \hilb{H}_a \otimes \hilb{H}_b $, $A_{a,b}$ to stress that $A$ belongs to $(\hilb{H}_{a,b})$, 
and, similarly, $\ket{\psi}_{a}$ and  $\Ket{A}_{a,b}$ to stress that $\ket \psi$ belongs to $\hilb H_a$ and $\Ket A$ belongs to $\hilb H_{a,b}$.

Using the Choi isomorphism \cite{choi},  we can associate   each completely positive map $\mathcal{T}_i : L (\hilb H_0) \to L(\hilb H_1)$ with a  positive operator $T_i \in L  (\hilb{H}_1 \otimes \hilb{H}_0 )$ given by 
\begin{align}\label{choi}
T_{i} = (\mathcal{T}_i \otimes \mathcal{I}_0 )(\KetBra{I}{I}_{0,0}) 
\end{align}
$\mathcal{I}_{0}$ being the identity map on $\hilb{H}_0$. 

In terms of the Choi operator, the condition that $ \map T$ is trace-preserving  (resp. trace-decreasing) becomes
\begin{align}\label{normforchan}
\Tr_{1} [T]  = I_{0}  \qquad  \left ( \mathrm{reps.}  \Tr_1 [T] \le I_0 \right)
\end{align}
where  $\Tr_1$ denotes partial trace over $\hilb{H}_1$. 
 
\begin{figure}[htbp]
\begin{center}
\includegraphics[width=\columnwidth ]{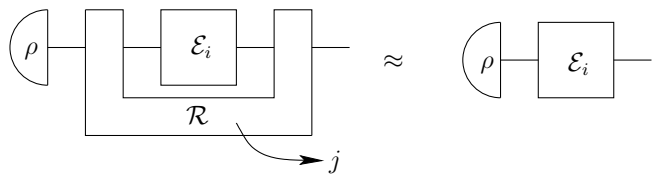}
\caption{Given a black box implementing an unknown channel $\mathcal E_i$ drawn from a set $\{ \map E_i \}$, we want to link it with
a quantum network $\mathcal{R}$ that
both gives an estimate $j$ of 
the parameter $i$ and affects the output state $\map E_i (\rho)$  as little as possible (here the symbol $\approx$ means that the network $\map R$ is optimized in such a way that the output of the two circuits on the left and on the right is as close as possible).
}
\label{combtradeoff}
\end{center}
\end{figure}

We now introduce the trade-off problem for quantum transformations.
Consider a quantum network $\mathcal R$ with an empty slot that can be linked with a variable quantum device, the input-output action of the latter being described by a channel in the set $\{\mathcal{E}_i\}$. Ideally, we would like the network $\mathcal{R}$ to give us some information about the channel $\mathcal{E}_i$ without affecting the output state $\mathcal{E}_i(\rho)$ that the channel should produce when an input state $\rho$ is fed in the corresponding device. 
(see Fig. \ref{combtradeoff}). However, as already mentioned, this is not possible in general. 

As we already mentioned, there are two extreme situations. On one extreme, if we are only interested in extracting information,   the best strategy is to apply the channel on one side of a suitable bipartite state $\sigma  \in L(\hilb H_o \otimes \hilb H_0  )$, thus getting the output state $(\map E_i \otimes \map I_0)(\sigma)$, and then to perform a suitable measurement $\{P_j\}$.  In this case the available use of the channel is consumed for estimation: after this step, the best we can do to produce an output state close to $\map E_i (\rho)$ is to apply to the input state $\rho$ some channel $\tilde {\map E}_j$ that depends on the outcome $j$ of our measurement.    On the opposite extreme, if we do not tolerate any disturbance,
the only possible  is to apply  the black box to the input state $\rho$. In this case we correctly obtain  the output state $\map E_i (\rho)$, but we have no measurement data to infer the identity of the unknown device.  
In the intermediate  cases, it is important to assess the maximum amount of information that can be gathered
without trespassing a given disturbance threshold. 
 
Since the trade-off problem involves optimization of quantum networks, we will use
the approach of \emph{quantum combs} developed in \cite{architecture, supermaps, comblong}.  This approach is based on the characterization of the most general transformations that quantum channels can undergo, and on realization theorems proving that all these abstract transformations can be implemented by quantum networks.  Since our theorems are constructive, this approach will also provide the explicit form of the optimal quantum network.

\section{Quantum combs and generalized instruments}\label{sec:combs}
By stretching and rearranging the internal wires,
 we can give to every quantum network the shape of a comb.
 The empty slots of the network
becomes the empty space between two teeth of the comb.

\begin{figure}[htbp]
\begin{center}
\includegraphics[width=\columnwidth ]{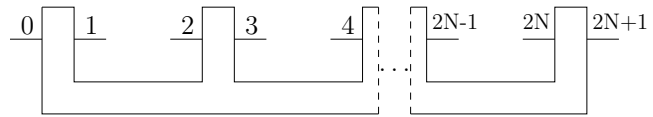}
\caption{A quantum comb with $N+1$ teeth. The flow of quantum information is from left to right. The input wires of the network are labelled with even numbers from 0 to $2N$, the output wires with odd numbers from 1 to $2N+1$.} 
\label{comb}
\end{center}
\end{figure}

 Referring to Fig. \ref{comb}, each wire
 is labeled with a natural number, which is even for the input wires and odd for the output ones; the
corresponding Hilbert spaces are
labelled accordingly.

 
If our network consists of a sequence of $N$ quantum channels (trace-preserving maps), then we call it \emph{deterministic}.  To every deterministic network we can associate a positive operator $R^{(N)} \ge 0$, called \emph{quantum comb}, satisfying the normalization condition \cite{watgut,architecture,comblong}
\begin{align} \label{normalization}
\Tr_{2k-1}[R^{(k)}] = I_{2k-2} \otimes R^{(k-1)} \qquad k = 1, \dots, N,
\end{align}
where $R^{(0)} =1$, $R^{(k)}
\in L(\sH_{\rm{out}_k} \otimes \sH_{\rm{in}_k})$
with
 $\sH_{\rm{in}_k} = \bigotimes_{n=0}^{k-1} \sH_{2n}$ and
$\sH_{\rm{out}_k} = \bigotimes_{n=0}^{k-1} \sH_{2n+1}$,
 is the comb of the reduced circuit obtained by 
discarding the last $N-k$ teeth.

The normalization condition of Eq. (\ref{normalization}) reflects the causal ordering in the deterministic network.   We will call a comb satisfying Eq. \eqref{normalization} \emph{deterministic}, and we will denote by  $\mathsf{DetComb}(\bigotimes_{i=0}^{2N-1}\hilb{H}_i)$ the set of all deterministic combs with the given ordering of the input and output spaces.  
A deterministic quantum comb with $N=1$  is simply the Choi operator of a quantum channel: in this case the condition of Eq. (\ref{normalization})  is equivalent to the normalization of the channel  given in Eq. (\ref{normforchan}).
Accordingly, $\mathsf{DetComb}( \hilb H_b  \otimes \hilb H_a )$  will be the set of (Choi operators of) quantum channels from $L(\hilb H_a)$  to $L(\hilb H_b)$.   
 
This framework of quantum combs can be easily extended to the case of networks consisting of quantum operations (trace-decreasing maps).
We call \emph{probabilistic comb} a positive operator $S^{(N)}\ge 0$ that is bounded by some deterministic comb, that is, an operator $S^{(N)}$ with the property
\begin{align}\label{probcombnormalization}
\exists R^{(N)} \in \mathsf{DetComb} \left(
  \bigotimes_{k=0}^{2N-1}\hilb{H}_k \right)   \mbox{ such that }   S^{(N)} \leq R^{(N)}
\end{align}
We will denote by $\mathsf{ProbComb}(\bigotimes_{k=0}^{2N-1}\hilb{H}_k)$ the set of all probabilistic combs with given ordering of the input and output spaces.  A probabilistic comb with $N=1$ is simply the Choi operator of a quantum operation: in this case the condition of Eq. (\ref{probcombnormalization}) is equivalent to the bound in Eq.  (\ref{normforchan}).  Accordingly,  $\mathsf{ProbComb}  (\hilb H_b \otimes \hilb H_a)$ will be the set of (Choi operators of) quantum operations from $L(\hilb H_a )$ to $L(\hilb H_b)$.   

Two quantum networks $\mathcal{R}^{(N)}$ and $\mathcal{S}^{(M)}$ can be linked 
together by connecting  some wires of $\mathcal{R}^{(N)}$
with some  wires of $\mathcal{S}^{(M}$.     Let us denote by $\mathsf J$ the set of wires that are connected and by $\mathsf K$ ($\mathsf L$) the set of wires of  $\map R^{(N)}$  ($\map S^{(M}$) that are not.  
The circuit resulting from the connection, denoted by 
$\mathcal{R}^{(N)} * \mathcal{S}^{(M)}$, has Choi operator  given by the \emph{link product}
\begin{align} \label{linkproduct}
R^{(N)} * S^{(M)} = \Tr_{\mathsf{J}}\left[  \left(R_{\mathsf{K, J} }^{(N)}  \otimes   I_{\mathsf  L}  \right) \left(I_{\mathsf K}  \otimes S^{ (M) T_{\mathsf J} }_{\mathsf{J,L}} \right)\right],
\end{align}
where $R^{(N)},S^{(M)}$ are the Choi operators of $\mathcal{R}^{(N)}$ and $\mathcal{S}^{(M)}$, respectively,
and  ${T_{\mathsf{J}}}$ denotes the partial transposition with respect to the  fixed orthonormal basis of Eq. (\ref{doubleketeq}).       
For example, let   $\mathcal{E}$ be a channel from $\hilb{H}_0$ to $\hilb{H}_1$ and $\mathcal{F}$ be a channel
from $\hilb{H}_1$ to $\hilb{H}_2$; then the Choi  operator of the composition $\mathcal{FE}$ is 
\begin{align}
 F*E = \Tr_{1}[ (F_{2,1}  \otimes I_0   )  (I_2 \otimes E_{1,0}^{T_{1}})].
 \end{align}

As a particular case for $\hilb H_0 \simeq \Cmplx$,   if  $\rho$ is a state on $\hilb H_1$ and $\map E$ is a channel from $\hilb H_1$ to $\hilb H_2$  one has 
\begin{equation}\label{action of channel}
\map E (\rho)  =  E_{21} * \rho_1  =  \Tr_1[ E_{21}  ( I_2 \otimes \rho^T_1) ] . 
\end{equation}


A deterministic (probabilistic) quantum comb, besides representing  a quantum network with some empty slots, can also represent  a quantum
channel (operation) $\mathcal{R}^{(N)}$ from $\hilb{H}_{even}: = \bigotimes_{k=0}^{N-1}  \hilb H_{2k}$ to $\hilb{H}_{odd}:=\bigotimes_{k=0}^{N-1} \hilb H_{2k+1}$. Due to the Choi isomorphism, the channel $\map R^{(N)}$ is in one-to-one correspondence with the comb $R^{(N)}$.  In the following two subsections we will exploit this correspondence to discuss  the physical realization of quantum combs, both in the deterministic case (subsec. \ref{subsect:detcombs}) and in the probabilistic case (subsec. \ref{subsect:probcombs}).

\subsection{Realization of deterministic combs}\label{subsect:detcombs}

The following theorem, proved in Ref.  \cite{algorithm}, gives an explicit construction for the realization of every deterministic quantum comb as a sequence of isometric channels. 
\begin{theorem}[Realization of deterministic combs] \label{dilationtheorem}
Every deterministic comb can be realized as a concatenation of
isometric channels in the following way:  

\begin{widetext}
\begin{equation}\label{combrealization}
  \begin{aligned}
 \Qcircuit @C=1em @R=.7em {
    & \qw\poloFantasmaCn{0} & \multigate{3}{\map R^{(N)}}  & \qw \poloFantasmaCn{1}&\qw \\
        & \qw\poloFantasmaCn{2} & \ghost{\map R^{(N)}}  & \qw \poloFantasmaCn{3}&\qw \\
    & \poloFantasmaCn{\vdots} & \pureghost{\map R^{(N)}}  &  \poloFantasmaCn{\vdots} &  \\
    & \qw\poloFantasmaCn{2N-2 } & \ghost{\map R^{(N)}}  & \qw
    \poloFantasmaCn{ 2N-1} &\qw} \end{aligned} 
=
  \begin{aligned}
\Qcircuit @C=1em @R=.7em  
{  & \qw \poloFantasmaCn{0} & \multigate{1}{\map V^{[1]}} & \qw \poloFantasmaCn{1} & \qw && \qw \poloFantasmaCn{2} & \multigate{1}{\map V^{[2]}} & \qw \poloFantasmaCn{3} & \qw  & \dots & & \qw \poloFantasmaCn{2N-2 } & \multigate{1}{\map V^{[N]}} &\qw \poloFantasmaCn{ 2N-1} &\qw \\ 
    &             & \pureghost{\map V^{[1]}} & \qw
    \poloFantasmaCn{A_1} & \qw & \qw & \qw &\ghost{\map V^{[2]}} & \qw
    \poloFantasmaCn {A_2} & \qw    & \dots&  &\qw
    \poloFantasmaCn{A_{N-1}} &\ghost{\map V^{[N]}} & \qw
    \poloFantasmaCn{A_N} &\measureD{I}} 
\end{aligned}
\quad ,
\end{equation}
\end{widetext}
where $A_k$ is an ancilla with Hilbert space $\hilb H_{A_i}$, $\map
V^{[k]}:  L(\hilb H_{2k-2} \otimes \hilb H_{A_{k-1}})$ is the channel
defined by $\map V^{[k]}  (\rho )  = V^{[k]}  \rho V^{[k]\dag},
\forall \rho \in L(\hilb H_{2k-2} \otimes \hilb H_{A_{k-1}})$ for a
suitable isometry $V^{[k]}:  \hilb H_{2k-2}  \otimes \hilb H_{A_{k-1}}
\to \hilb H_{2k-1} \otimes \hilb H_{A_k} $, and 
  represents the partial trace on $\hilb H_{A_N}$.  
Precisely, the ancillary Hilbert spaces $\hilb H_{A_{k-1}}, \hilb H_{A_k}$ are defined by $\hilb{H}_{A_0}:= \mathbb{C}$ and, for $k\ge 1$ and  $\hilb{H}_{A_k}: = \supp(R^{(k)*}) \subseteq \bigotimes_{n=0}^{2k-1}  \hilb H_n$, where $\supp(R^{(k)*})$ denotes  the support of the complex conjugate of $R^{(k)}$ in the fixed basis. 
   The isometry $V^{[k]}$  is given by the expression
\begin{widetext}
\begin{align}\label{structureofisometries}
\nonumber V^{[k]} : =& 
\left \{I_{2k-1}\otimes \left [\left( R^{(k)*}_{(2k-1)', \dots, 0' }  \right)^{\frac 12}  \left (I_{(2k-1)' ,(2k-2)'}  \otimes R^{(k-1)*}_{(2k-3)', \dots, 0'}\right)^{-\frac{1}{2}} \right]\right\} \quad \times \\  
&\qquad \times  \left(  \Ket{I}_{(2k-1),(2k-1)'}  \otimes  T_{(2k-2)' \leftarrow  (2k-2)}  \otimes I_{(2k-3)'}  \otimes \dots \otimes I_{0'}  \right),
\end{align}
\end{widetext}
where $\hilb H_{k'}  \simeq  \hilb H_{k}$  and   $T_{m \leftarrow n}$ is the teleportation operator from $\hilb H_n$ to $\hilb H_m$, given by $T_{m \leftarrow n}  = \sum_{k}  \ket k_{m}  \bra k_n $.


\end{theorem}

Note that the isometry $V^{[k]}$ defined in Eq. (\ref{structureofisometries}) has the correct  input and output Hilbert spaces. Indeed,  for $k=1$ one has $R^{(0)} =1$ and  the isometry $V^{[1]}$,  given by $V^{[1]} =  \left[ I_1 \otimes \left( R^{(1)*}_{1',0'}  \right)^{\frac 12}  \right]  (\Ket I_{1,1'}  \otimes T_{0 \to 0'} )$, sends vectors in $\hilb H_0$ to vectors in $\hilb H_1 \otimes  \hilb H_{A_1}$, where $\hilb H_{A_1} = \supp (R^{(1)*}_{1',0'})$ is  a subspace of $\hilb H_{1'}  \otimes \hilb H_{0'}$.   For $k>1$, since  $\hilb  H_{A_{k-1}}  =   \supp (R^{(k-1)*}_{(2i-3)', \dots, 0'})$ is a subspace of $\hilb H_{(2k-3)'} \otimes \dots \otimes \hilb H_{0'}$,  the isometry $V^{[k]}$  sends vectors in $\hilb H_{2k-2} \otimes \hilb H_{A_{k-1}}$ to vectors in $\hilb H_{2k-1} \otimes \hilb H_{A_k}$, as stated by the thesis.

\subsection{Generalized N-instruments}\label{subsect:probcombs}
In the discrete-outcome case,  a \emph{generalized N-instrument} is a set of probabilistic combs $\{ R_i\} \subset \mathsf{ProbComb}(\bigotimes_{k=0}^{2N-1}\hilb{H}_k)$ satisfying the normalization condition
\begin{equation}\label{geninstnorm} 
R^{(N)} :=  \sum_i R_i  \in \mathsf{DetComb}(\bigotimes_{k=0}^{2N-1}\hilb{H}_k).
\end{equation} 
When $N=1$ the notion of generalized instrument 
coincides with the usual notion of quantum instrument.  
Every $N$-instrument can be realized as a quantum network, due to an analogue of Ozawa's dilation theorem 
\cite{Ozawa}.  The proof of the dilation theorem for generalized $N$-instruments was originally presented in Ref. \cite{comblong}, and is combined here with Theorem \ref{dilationtheorem}. 

 \begin{theorem}[Realization of $N$-instruments]\label{realization-theorem}  Every generalized N-instrument $\{R^{(N)}_i\}\subset \mathsf{ProbComb}(\bigotimes_{k=0}^{2N-1}\hilb{H}_k)$ can be realized as a quantum network of isometric channels followed by a measurement one the last ancilla, as follows: 
Here $A_k$, $\hilb H_{A_k}$, and $V^{[k]}$ are the ancillas, the Hilbert spaces, and the isometries providing the realization of the deterministic comb $R^{(N)} = \sum_i R_i^{(N)}$, as given by  Theorem \ref{dilationtheorem}.     
represents the partial trace on $\hilb H_{A_N}$ with the operator $P_i$.   $\{P_i\}$ is a quantum measurement on the ancilla $A_N$, described by the POVM  
 \begin{equation}\label{POVMonancilla}
 P_i  =    \left(R^{(N)*}\right)^{- \frac 12} R^{(N)*}_i  \left(  R^{(N)*}\right)^{-\frac 12} .
 \end{equation}   
\end{theorem}

In our study of the information-disturbance trade-off we will  use generalized 2-instruments,  which can be graphically represented by  combs with two teeth and one empty slot where  the unknown black box can be inserted, as in Fig. \ref{combtradeoff}.   Since the value of $N$ is fixed to $N=2$, in the following we will drop the index $(N)$  in $R_i^{(N)}$ and $R^{(N)}$ and simply write $R_i$ and $R$.

\section{Information-disturbance tradeoff for unitary channels}\label{sec:tradeoff}

\subsection{Formulation of the problem}\label{subsec:formulation}

Suppose that  a black box performs an unknown unitary channel $\map U
(\rho) = U \rho U^\dag$, where the unitary $U \in \mathbb{SU} (d)$ is
randomly drawn according to the normalized Haar measure $\d U$.  Let
$\hilb H_1 \simeq \hilb H_2\simeq  \Cmplx^d$ be the input and output
Hilbert spaces for the unknown channel, respectively.   In order to
extract information we will then use a quantum network like that of
Eq. \eqref{geninstnorm} with $N=2$.        The network will be the  described by a generalized 2-instrument 
 $\{ R_{\hat U}\}$,  the outcome  $\hat U\in \mathbb{SU}(d)$   being the estimate of the unknown parameter $U$.   For every possible outcome $\hat U$, $R_{\hat U}$  is an element of $ \mathsf{ProbComb}(\bigotimes_{k=0}^{3}\hilb{H}_k) \}$, with $\hilb H_0 \simeq\hilb H_1 \simeq\hilb H_2 \simeq\hilb H_3 $, and one has the normalization condition
\begin{equation}\label{norm2inst}   
R := \int_{\mathbb{SU}  (d)}  \d \hat U    R_{\hat U}  \in  \mathsf{DetComb}  (\bigotimes_{k=0}^{3}\hilb{H}_k ), 
\end{equation}
which is the continuous version of Eq. (\ref{geninstnorm}).  Since there will be no ambiguity, from now on we will omit the domain of integration $\mathbb {SU} (d)$.  

When the unknown black box with channel $\map U$  is connected to the quantum network with generalized instrument $\{\map R_{\hat U}\}$, one obtains a set of quantum operations $\map R_{\hat U}  *  \map U$, each corresponding to a possible result of the measurement.   
However, to speak about the ``the probability of the outcome $\hat U$'' we need to know what input state $\rho$ is fed in the circuit: we cannot speak of the ``probability of a quantum operation" without specifying its input state.     If the input state is $\rho \in L(\hilb H_0)$,  then the probability is given by the trace of the output state $(\map R_{\hat U} * \map U)  (\rho)$:       
\begin{align}\label{probabilityofV}
p(\hat U|U,\rho)&=\Tr [ (\mathcal{R}_{\hat U} * \mathcal{U}) (\rho)] \nonumber \\
&=  \Tr [R_{\hat U}  (I_3 \otimes \Ket {U^*} \Bra {U^*}_{2,1}  \otimes  \rho_0^*) ],
\end{align}
where we used the link product of Eq. (\ref{linkproduct}) and Eq. (\ref{action of channel}) to compute the Choi operator of $\map R_{\hat U}*  \map U$ and the action of the channel $\map R_{\hat U} *  \map U$  on the state $\rho$, respectively. We also used the fact that $A^T  = A^*$  for every self-adjoint operator.  
 
To quantify the information gain and the disturbance we now introduce two suitable fidelities.   
Suppose that the black box performs the unitary channel $\map U$ and that the measurement outcome is $\hat U$.   In this case we quantify information gain with the fidelity 
\begin{equation}\label{pointwisegain}
g (\hat U, U)  =\frac 1 {d^2}  |\Tr[\hat U  U^\dag]  |^2 
\end{equation} 
Note that the maximum value of the fidelity is 1, and is achieved if and only if $\hat U =\omega U$ for some phase $|\omega| = 1$, that is, if and only if the two unitary channels $\hat {\map U}$ and $\map U$ coincide.  The fidelity $g(\hat U, U)$ enjoys the invariance property  
\begin{equation}\label{invfid}
g(\hat U, U) =  g(  V  \hat U  W, V  U W) \qquad \forall V, W  \in \mathbb{SU} (d).
\end{equation}
Averaging  the fidelity with the probability of the estimate $\hat U$ given the true value $U$ and the input state $\rho$, we then obtain the \emph{average information gain} 
\begin{align}  
G_{\rho}  &:=  \int  \d  U \int   \d \hat U    p(\hat U|  U, \rho)   g (\hat U,U).
\end{align}   
In our analysis we will always assume that the input state is given by $\rho  =  I/ d$.  The reason for this choice is that the condition $\rho =  I/d$ arises in two relevant scenarios: 
\begin{enumerate}
\item when the input system (wire $0$) of the circuit is prepared  in a maximally entangled state with some reference system $0'$. This is the case in the cryptographic protocols of Refs. \cite{pingpong,clonunit}  (and, in the infinite energy limit, also in in the continuous variable scenario of  Ref. \cite{pirandola} ).               
\item when the input system is prepared at random in one of the states of an ensemble $\{\rho_i, p_i\}$, with the property that $\sum_{i}  p_i \rho_i  = I/d$. This is the case of the cryptographic protocol of Ref. \cite{mancini}.     
\end{enumerate}

Since we are setting $\rho = I/d$, we will drop the subscript $\rho$  in $G_\rho$.  Using Eqs. (\ref{probabilityofV})  and  (\ref{pointwisegain}), the expression for the information gain $G$  is    
\begin{align}\label{gain}
G  =\frac 1{d^3} \int \!\! \d U \! \int \d \hat U   \Tr_{3,0}[\Bra{U^*}_{2,1} R_{\hat U} \Ket{U^*}_{2,1}] |\BraKet{\hat U}{U}|^2 .
\end{align}

We now introduce our figure of merit for the minimization of the disturbance.   To this purpose, we consider the  \emph{channel fidelity}  \cite{raginski} between the overall quantum operation $\mathcal{R}_{\hat U} * \mathcal{U}$ performed by the network
 and the input channel $\mathcal{U}$.   This is the fidelity between the two output states produced by the two operations $\map R_{\hat U} * \map U$ and $\map U$ when applied on one side of the maximally entangled  state  $\Ket \Phi  =  \frac 1 {\sqrt d}   \Ket I_{0,0'}$. In terms of Choi operators, the channel fidelity  is given by 
 \begin{align*}
 F (\map R_{\hat U}  * \map U, \map U) & =  \frac 1 {d^2}   \Bra {U}_{3,0}    \left(   R_{\hat U} * \KetBra  U U_{2,1}  \right)   \Ket{U}_{3,0} \nonumber  \\
  & =   \frac 1 {d^2}   \Bra {U}_{3,0}     \Bra{U^*}_{2,1}   R_{\hat U}  \Ket  U_{3,0} \Ket {U^*}_{2,1}  ,
 \end{align*}
 where we used the fact that, by definition of Choi operator (Eq. (\ref{choi})),   one has   $ (\map E \otimes \map I) (\Ket \Phi \Bra \Phi) =  E/d$ for every quantum operation $\map E$.

 Averaging over the outcomes and  the true values we then obtain the \emph{average fidelity}  
 \begin{align}\label{avefid}
F := & \frac 1 {d^2} \int  \d U  \int \d \hat U     \Bra {U}_{3,0}     \Bra{U^*}_{2,1}   R_{\hat U}  \Ket  U_{3,0} \Ket {U^*}_{2,1} \nonumber \\
 = &    \frac{1}{d^2} \int  \! \! \d U   \Bra{U}_{3,0}\Bra{U^*}_{2,1} R  \Ket{U}_{3,0}\Ket{U^*}_{2,1}.
\end{align}

Note that the fidelity $F$ naturally arises also in the case where the input state at the wire $0$ is a pure state $\varphi = \ket \varphi \bra \varphi $  chosen at random according to the uniform measure on pure states: in this case the fidelity between  $\map R_{\hat U}  * \map U  (\varphi)$ and $\map U(\varphi)$, averaged over $\varphi, U$, and $\hat U$,  is given by
\begin{align*}
F'  & =  \int \d \varphi \int \d  U  \int \d \hat U     \Tr[ (U \varphi U^\dag)   \left(   \map R_{\hat U}  *  \map U\right) (\varphi) ] \\
&= \int \d \varphi\int \d U \int \d \hat U       \Tr[\left ( R_{\hat U} *     \KetBra  U U_{2,1} \right)  ( U \varphi  U^\dag  \otimes  \varphi^*)  ] \nonumber \\
&=  \int  \d \varphi \int \d U    \bra  { U  \varphi}_3   \Bra  {U^*}_{2,1} \bra{\varphi^*}_0   R     \ket{  U \varphi}_3   \Ket  {U^*}_{2,1} \ket {\varphi^*}_0 \nonumber \\
& = \int \d U  \left(  \frac d {d+1}    F(\map R *  \map U, \map U ) \right.  + \nonumber \\
& \qquad +  \left. \frac 1 {d(d+1)} \Tr[ \Bra{U^*}_{2,1}  R  \Ket { U^*}_{2,1} ] \right)  \nonumber \\
&=     \frac d {d+1}  F  + \frac{1}{d+1} , \nonumber 
\end{align*}     
having used the relation 
\begin{equation} 
\int \d \varphi     \varphi \otimes \varphi^*  =   \frac {   \Ket I \Bra I  +  I\otimes I } {d(d+1)} 
\end{equation}
and the normalization $Tr [R]  =d^2$, which follows directly from Eq. (\ref{normalization})  in the $N= 2$ case with $\hilb H_3 \simeq  \hilb H_2 \simeq \hilb H_2 \simeq \hilb H_0 \simeq \Cmplx^d$.  
Since there is a trade-off,  the information gain $G$ and the fidelity $F$ cannot achieve their maximum values at the same time.  Therefore, we will introduce a weight $0\le p\le 1$ that quantifies how much we care about information extraction versus disturbance minimization, and our figure of merit will be  the convex combination  $p G +(1-p) F$.   The extreme case $p=0$ and (resp. $p=1$) corresponds to the situation where we do not  tolerate any disturbance (resp. where we are only interested in extracting maximum information). The trade-off curve obtained by the maximization of $p G + (1-p)  F$ for all possible values of $p \in [0,1]$ is the same curve that would  be obtained by maximizing $F$ for given $G$  (i.e. by finding the minimum disturbance for given amount of extracted information), or by maximizing $G$ for given $F$ (i.e. by finding the maximum amount of extractable information for given disturbance threshold).      

\subsection{Symmetry of the estimating network}\label{subsec:symmetries}

Here we exploit the symmetries of the figure of merit $p G + (1-p)F$ to simplify the optimization problem.  The crucial simplification comes from the following Theorem, which states the symmetry properties of the optimal generalized instrument:  
\begin{theorem}[Symmetries of the optimal instrument]\label{theo:covariance}
Let $G$ and $F$  be the information gain and the fidelity defined in Eqs. (\ref{gain}) and (\ref{avefid}). For every $p \in [0,1]$, the generalized instrument that maximizes $pG + (1-p) F$  can be chosen to be \emph{covariant}, that is, of the form
\begin{align}\label{def R_V}
R_{\hat U} = (\hat {\map U}_{3} \otimes \hat {\map U}_{2}^* \otimes \map I_{1,0}) (\Xi) ,
\end{align}
for some positive operator $\Xi \in L (\hilb H_3\otimes \hilb H_2 \otimes \hilb H_1 \otimes \hilb H_0)$. 

Moreover, the operator  $\Xi $ satisfies the commutation relation
\begin{equation}\label{commrel}
[\Xi, V_3  \otimes  V^*_2 \otimes  V_1\otimes V^*_0 ] = 0 \qquad \forall V \in \mathbb{SU} (d).
\end{equation}        
\end{theorem}

\Proof  The proof is based on the same argument used for the  proof of Lemma 2 in Ref. \cite{unitaryestimation}.   Consider an arbitrary generalized instrument $\{  R_{\hat U}\}$.  Using the invariance of the Haar measure and of the fidelity $g(\hat U, U)$  (Eq. (\ref{invfid})),  it is easy to check that the values of $F$ and $G$ in   Eqs. (\ref{gain}) and (\ref{avefid}) do not change if each $R_{\hat U}$ is replaced by the group average  
\begin{align}\label{tritatutto} 
R_{\hat U}'  := \int  \d V \d W    (\map V_{3} \otimes \map V_2^* \otimes \map W_1 \otimes \map W^*_0) (R_{V^\dag  \hat U  W}) ,
\end{align} 
where $\map V$ ,$\map V^*, \map W  ,\map W^*$ are the unitary channels corresponding to the unitaries $V, V^*,W, W^*$, respectively.  Note that $\{ R_{\hat U}'  \}$ is still a generalized instrument, because it satisfies the normalization  condition  of Eq. (\ref{norm2inst}). Moreover,  from Eq. (\ref{tritatutto}) is is clear that  $\Xi :=  R'_I$ satisfies the commutation relation of Eq. (\ref{commrel}).    Finally, from Eq. (\ref{tritatutto}) it is also clear that for every   $ \hat U, V, W\in \mathbb{SU} (d)$ one has  
\begin{align*}
R'_{V\hat U W^\dag}  =   (\map V_{3} \otimes \map V_2^* \otimes \map W_1 \otimes \map W^*_0) (R'_{\hat U })  
\end{align*} 
Taking $\hat U= W =I$ one then obtains $R'_{V} = (\hat {\map V}_{3} \otimes \hat {\map V}_{2}^* \otimes \map I_{1,0}) (\Xi)$, namely $R_{\hat U}'$ is of the form of Eq. (\ref{def R_V}).  Since the substitution $\{R_{\hat U}\} \longrightarrow \{R_{\hat U}'\}$ can be done for every generalized instrument, in particular it can be done for the optimal one. \qed

Using Theorem \ref{theo:covariance}, we can now express the normalization condition of Eq. (\ref{norm2inst})  in a particularly simple way. 
Indeed, Eqs. (\ref{def R_V})  and (\ref{commrel})  imply that the normalization operator  $R = \int  \! \! \d \hat U R_{\hat U}$ satisfies the commutation relation
\begin{equation*}
[R, V_{3} \otimes V_{2}^* \otimes W_{1} \otimes W_{0}^* ] = 0  \qquad \forall V,W \in \mathbb{SU} (d).
\end{equation*}
The Schur lemma then implies  $\Tr_3 [R]  = I_2 \otimes R^{(1)}$ for some positive operator $R^{(1)} \in L(\hilb H_1 \otimes \hilb H_0)$, and  $\Tr_{1}  [R^{(1)}] =  \alpha I_0   $ for some positive number $\alpha \in \mathbb R $.     Therefore, the normalization condition for $R$ to be a deterministic comb  (Eq. (\ref{normalization}) for $N=2$)  becomes 
 trivially $\Tr [R]  = d^2$, or, equivalently,  
\begin{align}\label{normalization2}
\Tr[\Xi]= d^2.
\end{align}

\subsection{Optimal trade-off curve}\label{subsec:curve}

Exploiting Theorem \ref{theo:covariance} and the Schur's lemmas we can now rewrite the figure of merit as:
\begin{align}\label{eq:convexcomb}
p G + (1-p) F = \Tr[(  p \Lambda_G   +  (1-p)  \Lambda_F  ) \Xi ]
\end{align}
where $\Lambda_G$ and $\Lambda_F$ are  the positive operators given by
\begin{align}
  \Lambda_F &:=  \frac{1}{d^2(d^2-1)} [I_{3,2,1,0} + d^2 P_{3,2} \otimes P_{1,0} -  \nonumber \\
&\qquad - P_{3,2} \otimes I_{1,0}- I_{3,2} \otimes P_{1,0}] \nonumber \\
 \Lambda_G &:=  \frac 1 d    (I_3   \otimes \Tr_{3,0} [\Lambda_F]   \otimes I_0)\\
  &= \frac{1}{d^2(d^2-1)}
 \left[ 
 \left( 
 1-\frac{2}{d^2} 
 \right) 
 I_{3,2,1,0}  + 
 \right.  \nonumber \\
 & \left.  \qquad+  I_{3} \otimes P_{2,1}  \otimes I_0\right] 
\end{align}
$P=d^{-1}\KetBra{I}{I}$ being the projector on the one-dimensional invariant
subspace of $V\otimes V^*$.

Since the only restrictions on $\Xi$ are positivity and the normalization given by Eq. (\ref{normalization2}), 
the optimal choice is to take $\Xi$ proportional to the projector on the eigenvector of
 $p\Lambda_G+(1-p)\Lambda_F$ corresponding to the maximum eigenvalue;  up to normalization,  this eigenvector the  can be shown  to be of the form \cite{ent-tradeoff}
\begin{align}\label{chi}
 \ket{\chi} = x \Ket{I}_{3,0}\Ket{I}_{2,1} + y \Ket{I}_{3,2}\Ket{I}_{1,0} ,
\qquad x,y \in \mathbb{R}^+
\end{align}
In order to satisfy Eq. (\ref{normalization2}), we then  choose $\Xi =  \ket \chi \bra \chi$ with the normalization $\braket \chi \chi   = d^2$.   Reminding Eq. (\ref{def R_V}) we get
\begin{align}\label{chiu}
&R_{\hat U}=\ketbra{\chi_{\hat U}}{\chi_{\hat U}}  \nonumber \\
 &\ket{\chi_{\hat U}} := x \Ket{\hat U}_{3,0}\Ket{\hat U^*}_{2,1} + y \Ket{I}_{3,2}\Ket{I}_{1,0}.
\end{align}
The normalization of $\chi$ implies that $x$ and $y$  obey the quadratic equation
\begin{align}\label{xandy}
x^2+y^2+ \frac {2xy} d =1 
\end{align}
Note that in the above equation there is just one free parameter (either $x$ or $y$), which can be expressed e.g. as  a function on the trade-off ratio $p$.
Fidelity and gain can be calculated in terms of the parameters $x$ and $y$, thus getting the following expressions
\begin{align}\label{FGxy}
F  = 1- \frac{d^2-2}{d^2}x^2 \qquad G = \frac{2-y^2}{d^2}
\end{align}
The extreme situation of minimum disturbance (resp. maximum extraction of information) can be retrieved in the extreme case $x=0,y=1$  (resp. $y=1, x=0$). 
     Indeed, when $x=0, y=1$, one has
$R_{\hat U} = \Ket{I}\Bra{I}_{3,2} \otimes \Ket{I}\Bra{I}_{1,0}$ for all $\hat U$. In this case, there is no information extracted and one has  $ \map R_{\hat U}  * \map U = \map U$,  that is,  the instrument realizes the identity map.   Accordingly, the  fidelity $F$ reaches its maximum
 $F=1$, while the  information gain 
takes its minimum  $G = \frac{1}{d^2}$, the value achieved by a random guess according to the Haar measure.
In the opposite case $x=1, y=0$ one has instead  $R_{\hat U}  =   \Ket {\hat U}\Bra{\hat U}_{3,0}  \otimes \Ket{\hat U^*}  \Bra {\hat U^*}_{2,1} $, which implies  $\map R_{\hat U} *  \map U =  |\Tr[  \hat U  U^\dag]|^2   \hat  {\map U}$.  This means that in this case our circuit performs the optimal estimation of $U$ \cite{unitaryestimation}, and then  executes the transformation $\hat{\map  U}$ on the input state.    Accordingly,   the fidelity drops to its minimum value 
 $F=\frac{2}{d^2}$ and the information gain reaches its maximum  $G = \frac{2}{d^2}$.   

Following Ref. \cite{ent-tradeoff}  we  introduce now the information
variable $0\le I \le 
1$  and the disturbance variable $ 0\le  D \le 1$, given by  
\begin{eqnarray}
I:=\frac{G - G_{min}}{G_{max}-G_{min}} \qquad D:= \frac{F_{max}-F}{F_{max}-F_{min}} \nonumber
\end{eqnarray}
where $G_{max}=2/d^{2}$, $G_{min}=1/d^{2}$, $F_{min}= 2/d^{2}$ and $F_{max}=1$.  Note that $I=0$  ($D=0$) corresponds to no information  (no disturbance) and $I=1$ ($D=1$) corresponds to maximum information (maximum disturbance).  
  
Using Eq. (\ref{FGxy})  and the definitions of $I$ and $D$
 it is immediate to obtain the relation
\begin{align}\label{xyID}
x = \sqrt D \qquad y = \sqrt{1-I}.
\end{align}
Substituting the above  equation in the normalization condition of Eq. (\ref{xandy}) we finally obtain the curve of the optimal trade-off:  
\begin{eqnarray}\label{tradeoff2}
d^2(D-I)^2 -4D (1-I)=0,
\end{eqnarray}
\begin{figure}[htbp]
\begin{center}
The corresponding plot  is showed in Fig. \ref{fig1}. 
\includegraphics[width=\columnwidth ]{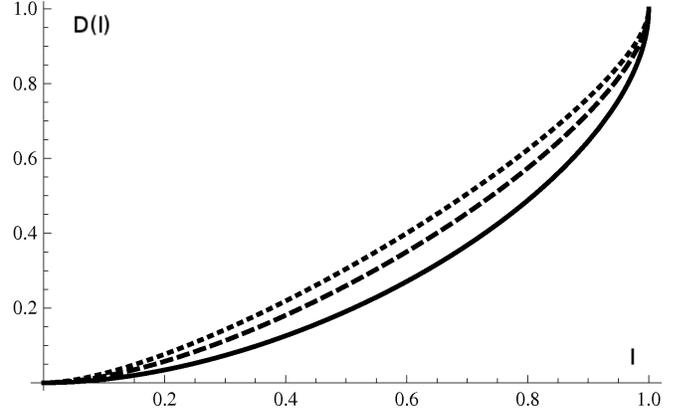}
\caption{Plot of the minimum disturbance $D(I)$ as a function of the information $I$  (Eq. (\ref{tradeoff2})),  for various values of $d$: $d=2$ solid line,  $d=3$ dashed line,  $d=4$ dotted line}
\label{fig1}
\end{center}
\end{figure}

\subsection{Optimal quantum network}\label{subsec:network}

We now use Theorems   \ref{dilationtheorem} and \ref{realization-theorem}  to construct explicitly the optimal network achieving the trade-off of Fig. \ref{fig1}.  
The optimal network will be derived for every possible value of the parameters $x,y$ with $0\le x,y \le 1$ belonging to the curve $  x^2 +  y^2  +  2 xy/d  =1 $. 
Since $x$ and $y$ can be easily expressed in terms of the information $I$  and the disturbance $D$ (Eq. (\ref{xyID})), our construction provides the optimal network for very point in the optimal trade-off curve depicted in Fig. \ref{fig1}. 

According to Theorem \ref{realization-theorem} the generalized instrument $\{R_{\hat U}\}$ is implemented as a sequence of $N=2$ isometries  $V^{[1]}: \hilb H_0  \to \hilb H_1 \otimes \hilb H_{A_1} $ and $V^{[2]}  : \hilb H_2 \otimes \hilb H_{A_1}  \to \hilb H_3 \otimes \hilb H_{A_2}$, followed by a measurement $\{P_{\hat U}\}$ on the ancilla  $A_2$.  The ancillary Hilbert space  $\hilb H_{A_1}$  ($\hilb H_{A_2}$) is given by $\hilb H_{A_1}  =  \supp ( R_{1',0'}^{(1)*} )  \subseteq  \hilb H_{1'} \otimes \hilb H_{0'}$ ($\hilb H_{A_2}  =  \supp ( R_{3',2',1',0'}^*)  \subseteq  \hilb H_{3'}  \otimes \hilb H_{2'} \otimes \hilb H_{1'} \otimes \hilb H_{0'}$ ), with $\hilb H_{k'}  \simeq \hilb H_k$ 
The isometries $V^{[1]},V^{[2]}$ are obtained from the realization of the deterministic comb $R =  \int \d \hat U  R_{\hat U}$ (Theorem \ref{dilationtheorem}) and the  ancilla measurement is given by the POVM 
\begin{align}\label{povmdacitare}
P_{\hat U}  =    \left (R^* \right)^{-\frac 12 *}  R^*_{\hat U}  \left(R^* \right)^{- \frac 12} .
\end{align}  

Let us start from the construction of the isometries.  By explicit calculation we find 
\begin{align}
R  &=  \left (   \frac{x^2 d^2}  {d^2-1}  +  y^2 d^2   + 2 xy d  \right)    (P_{3,2}  \otimes  P_{1,0}  )+ \nonumber\\
& \qquad + \left( \frac{x^2}  {d^2 -1} \right) (I_{3,2,1,0}  -  P_{3,2}  \otimes I_{1,0}  -  I_{3,2}  \otimes P_{1,0})  . \nonumber 
\end{align}   
Taking the partial trace on $\hilb H_3$ and using the condition $\Tr_{3} [ R]  =  I_2 \otimes R^{ (1) }$ we then obtain
\begin{align}
R^{(1)}  &=  \frac{(x+ d y)^2}{ d}   P_{1,0}  +  \frac{x^2}  d  (I - P)_{1,0},  \nonumber
\end{align}
and, therefore, 
\begin{align}
\left(R^{(1) *}\right)^{\frac 12}  &=  \frac 1 {\sqrt d}   \left(  y d   P_{1,0}    +  x  I_{1,0} \right) \\
\left ( R^{(1)  *}   \right)^{-\frac 12}  & = \sqrt d  \left ( \frac {- y d }{ x(x+ y d)}   P_{1,0}  +  \frac 1 x  I_{1,0}  \right)  \label{R1-12}        
\end{align}
According to Eq. (\ref{structureofisometries}), the isometry  $V^{[1]}  :  \hilb H_0  \to  \hilb H_1  \otimes  \hilb  H_{A_1}  \subseteq  \hilb H_{1,1',0'}$  is given by 
\begin{align}
V^{[1]} &= \left [  I_1 \otimes \left(  R^{(1) *}_{1',0'}  \right)^{\frac 12}  \right]  (\Ket{I}_{1,1'}\otimes T_{0' \leftarrow 0} )  \nonumber \\
&=   \frac 1 {\sqrt d}\left(  y  T_{1 \leftarrow 0} \otimes \Ket{I}_{1',0'}  
+ x   \Ket{I}_{1,1'}\otimes T_{0' \leftarrow 0}\right) \nonumber
\end{align}
If we input a pure state $\ket{\psi} \in \hilb H_0$  the output is then the superposition 
\begin{equation*}
V^{[1]} \ket \psi_0  = \frac{ y}{\sqrt d} \ket{\psi}_{1} \Ket{I}_{1',0'}  
+ \frac x  {\sqrt d}  \Ket{I}_{1,1'} \ket{\psi}_{0'}.
\end{equation*}
Intuitively, we can understand the action of $V_1$  as a superposition of two different processes: 
\begin{enumerate}
\item with amplitude $y $ the quantum state $\ket \psi_0$  is propagated undisturbed from the input system $0$ to the output system $1$, so that the unknown unitary $U$ can act on it.   As we will see in the following, the maximally entangled state $\Ket {\Phi}_{1',0'}  = \frac 1 {\sqrt d}  \Ket{I}_{1',0'}$ will then serve as a resource to teleport the state $U\ket \psi_2$ to the output node $3$. 
\item with amplitude $x$ the state $\ket \psi_0$ is  transferred to the ancillary degree of freedom $0'$: in this case the  unknown unitary will not act on it, but, instead, it will act on the maximally entangled state $\Ket \Phi_{1,1'}  = \frac1{\sqrt d}  \Ket I_{1,1'}$,  thus producing the state $\Ket {\Phi_U}_{2,1'}  =\frac 1 { \sqrt d}  \ket U_{2,1'}$. As we will see in the following, the state $\Ket {\Phi_U}_{2,1'}$ will be used for the optimal estimation of $U$.  Finally, the state $\ket \psi_{0'}$ will be transferred to the output system $3$, and,   depending on the estimate, a transformation $\hat U$  will be applied on it.     
\end{enumerate}
       
The  isometry $V^{[2]}  \hilb H_{2,1',0'}   \supseteq  \hilb H_2 \otimes  \hilb H_{A_1}  \to  \hilb H_{3} \otimes \hilb H_{A_2}  \subseteq \hilb H_{3,3',2',1',0'}$ is given by:
\begin{align}\label{v2}
V^{[2}] &= \left\{  I_3 \otimes \left[  \left(   R^*_{3',2',1',0'}  \right)^{\frac 12}  \left (  I_{3',2'}  \otimes  R^{ (1) *}_{1',0'}  \right)^{- \frac 12}  \right]  \right\}  \times  \nonumber \\
 & \qquad  \times  ( \Ket{I}_{3,3'} \otimes T_{2' \leftarrow 2}  \otimes I_{1',0'})  
\end{align}   
On the other hand, using Eqs. (\ref{povmdacitare}) and (\ref{chiu})  the POVM $\{P_{\hat U}\}$ on $\hilb H_{A_2}$ can be written as
\begin{align}\label{pu}
P_{\hat U} = \ketbra{\eta_{\hat U} } {\eta_{\hat U} } \qquad
\ket{\eta_{\hat U} } := \left( R^* \right)^{-\frac12} \ket{\chi^*_{\hat U} } 
\end{align}
Combining the isometry $V^{[2]}$  with the POVM $\{P_{\hat  U}\}$ we then obtain the instrument $\{\map T_{\hat U}\}$, with $\map T_{\hat U}  : L ( \hilb H_2 \otimes  \hilb H_{A_2} )  \to L(\hilb H_3)$ given by  
\begin{equation}
\mathcal{T}_{\hat U}  (\rho) =    \Tr_{A_2}  [  V^{[2]}   \rho  V^{[2]\dag}  (I_3 \otimes P_{\hat U}) ], \qquad \forall \rho \in L (\hilb H_2 \otimes  \hilb H_{A_1}) \nonumber
\end{equation}
We now use Eqs. (\ref{v2})  and (\ref{pu}) to show that the instrument $\{\map T_{\hat U}\}$ has a very simple  form. 
To construct $\map T_{\hat U}$ explicitly we start from the Kraus form $\map T_{\hat U}  (\rho)  =  K_{\hat U}  \rho  K_{\hat U}^\dag $, where the Kraus operator $K_{\hat U}$ is given by       
\begin{align*}
K_{\hat U}  &=    (I_3 \otimes \bra  {\eta_{\hat U}}_{3',2',1',0'}  )  V^{[2]} \\
&  =  \left ( I_3 \otimes  \bra{ \chi_{\hat U}^* }_{3',2',1',0'}    \right)  \left(   I_3 \otimes  I_{3',2'}  \otimes R^{(1)*}_{1',0'}  \right)^{-\frac 1 2} \times \\ 
& \qquad  \times     \left(  \Ket I_{3,3'}  \otimes T_{2' \leftarrow 2}  \otimes I_{1',0'}\right)    ,
\end{align*}
having used Eqs. (\ref{v2}) and (\ref{pu}).    Now, from Eq. (\ref{chiu}) we have $\bra {\chi^*_{\hat U}}_{3',2',1',0'} =  \bra \chi_{3',2',1',0'}  \left( \hat U^T_{3'}  \otimes \hat U^\dag_{2'} \otimes I_{1',0'}   \right) $, and, therefore
\begin{align}\nonumber
K_{\hat U}  &    =  \left ( I_3 \otimes  \bra{ \chi}_{3',2',1',0'}    \right)  \left(   I_3 \otimes  I_{3',2'}  \otimes R^{(1)*}_{1',0'}  \right)^{-\frac 1 2}  \nonumber \times \\ 
& \qquad  \times     \left(  \Ket {\hat U}_{3,3'}  \otimes   \hat U^\dag_{2'} T_{2' \leftarrow 2}  \otimes I_{1',0'}\right)  \nonumber  ,
\end{align}
Finally, inserting Eqs. (\ref{chi}) and (\ref{R1-12}) in the above expression we obtain
\begin{align}\nonumber
K_{\hat U} = \sqrt{d}  \left (   \Bra{\hat U}_{2,1'} \otimes  \hat U_3  T_{3 \leftarrow 0'}   \right),
\end{align}
which implies that the instrument $\{\map T_{\hat U}\}$ is given by 
\begin{equation}\label{instrument}
\map T_{\hat U}  (\rho)  =   \hat U_3  T_{ 3 \leftarrow   0'}  \left (  d  \Bra {\hat U}_{2,1'}  \rho   \Ket {\hat U}_{2,1'}  \right)  T_{0'  \rightarrow 3} \hat U_3^\dag,
\end{equation}
where we introduced the redundant notation $T_{0' \rightarrow 3}  :=  T_{3 \leftarrow  0' }^\dag  =\sum_{n=1}^d    \ket n_{0'}  \bra n_3   \equiv  T_{0' \leftarrow 3}$, to make the expression clearer.

  The interpretation of Eq. (\ref{instrument})
 is straightforward:  to implement the instrument $\{\map T_{\hat U}\}$  we only have to perform on system $2$ and on the ancila $1'$  the Bell measurement $\{  B_{\hat U}  \}$ with POVM $ B_{\hat U} = d   \Ket {\hat U}  \Bra{\hat U} $, and, depending on the outcome, to perform the unitary $\hat U$ on the output system $3$, which is obtained from the ancilla $0'$ just by relabelling (represented here by the teleportation operator $T_{3 \leftarrow 0'}$).   In other workds, the instrument $\{ \map T_{\hat U} \}$ is just obtained by a Bell measurement followed by unitary feed-forward.  
Remarkably,  $\{\map T_{\hat U}\}$ is independent of the trade-off parameters $x,y$: this means that, after we performed the isometry $V^{[1]}$, the remaining part of the optimal network is independent of the particular value of the information-disturbance rate.  The reason for this is that the combination of Bell measurement and feed-forward realized by the instrument $\{\map T_{\hat U}\}$ can work both as a teleportation protocol (in the case of no-disturbance)  and as an estimate-and-reprepare strategy (in the case of maximal information extraction).

\begin{figure}[htbp]
\begin{center}
\includegraphics[width=\columnwidth ]{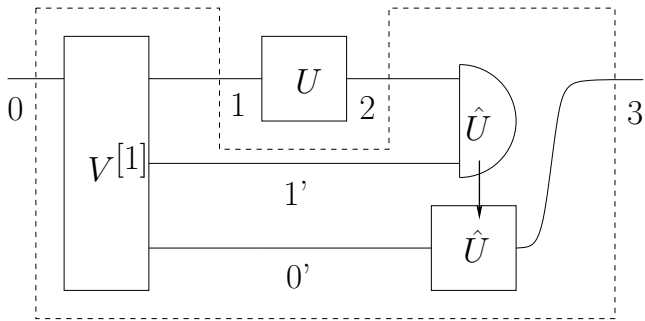}
\caption{Optimal quantum network for the information-disturbance trade-off.
The input state $\ket \psi $ enters from the wire $0$.   Then,  the isometry $V^{[1]}$ prepares a coherent superposition $\frac{1}{\sqrt{d}}\left( y  \ket \psi_1 \Ket{I}_{1',0'}   
+ x\Ket{I}_{1,1'} \ket{\psi}_{0'} \right)$ which is tuned by the parameters $x= \sqrt D $ and $y = \sqrt{1-I}$, whose value depend on the information-disturbance rate.
After that, the unknown unitary $U$ is applied between the nodes $1$ and $2$ of the network.  Finally, a Bell measurement is performed and, depending on the result, the unitary transformation $\hat U$  is performed on the output system $3$.}
\label{fig4}
\end{center}
\end{figure}

To summarize the results of this section, we give now the step-by-step evolution of a pure state $|\psi\>_0$ in the optimal quantum network:
\begin{align*}
\ket{\psi}_0 
&\xrightarrow{ V^{[1]}  }  
\frac{1}{\sqrt{d}}\left( y  \ket{\psi}_{1} \Ket{I}_{1',0'} 
+ x\Ket{I}_{1,1'}\ket{\psi}_{0'}\right)   \\
&\xrightarrow{ U   }
\frac{1}{\sqrt{d}} \left( y  U\ket{\psi}_{2}
  \Ket{I}_{1',0'} + x\Ket{U}_{2,1'} \ket{\psi}_{0'}\right)\\
&\xrightarrow{ \sqrt d  \Bra  {\hat U}}
y  {\hat U}^\dag U \ket{\psi}_{3} + x \Tr[\hat U U^\dag]  \ket{\psi}_{3}\\
&\xrightarrow{ \hat U}
y  U \ket{\psi} _{3} + x \Tr[\hat U U^\dag]  \hat U \ket{\psi}_{3}.
\end{align*}

The action of the whole network is depicted in Fig. \ref{fig4}.

\section{Conclusions}\label{sec:conclusions}

In this work we addressed the fundamental problem of  the information-disturbance trade-off in the estimation of an unknown quantum transformation. In particular, we completely solved the problem in the case of a unitary transformation, randomly distributed according to the Haar measure.

Interestingly, the analytical expression of the optimal trade-off curve given in Eq. (\ref{tradeoff2})) happens to coincide with the trade-off curve for the estimation of a maximally entangled state  \cite{ent-tradeoff}.  Note, however, that this is not a trivial consequence of the Choi isomorphism  $U  \rightarrow  \frac 1 {\sqrt d} \ket U$:  while this mathematical correspondence is one-to-one, operationally it cannot be inverted \emph{with unit probability}.  In other words, once the transformation $U$ has been applied to the maximally entangled state $\frac 1  {\sqrt d} \Ket I$, it is irreversibly degraded, and can be retrieved only probabilistically.  For this reason, there is no operational relation between the information-disturbance trade-off for unitary transformations and that for maximally entangled states  (none of them is a primitive for the other).  Indeed, the optimal quantum network for unitary transformations depicted in Fig. \ref{fig4}  is quite different from the optimal network for  maximally entangled states.   In our case the optimal network consists of \emph{i)} a first interaction that produces a quantum superposition with amplitudes depending on the desired information-disturbance rate and \emph{ii)} of a Bell measurement followed by unitary feed-forward.  

Besides its fundamental relevance, the information-disturbance trade-off for transformations is also interesting as a possible eavesdropping strategy in cryptographic protocols where the secret key is encoded in a set of unitary transformations, as it happens in the two-way protocols of Refs.  \cite{pingpong,mancini,clonunit,pirandola}.  Notice, however, that for protocols where the secret key is encoded in a set of \emph{orthogonal} unitaries,  like those  of Refs. \cite{pingpong,mancini,pirandola}, the security of the protocol is not based on the information-disturbance trade-off  studied in this paper.  Indeed, since  the unitaries are orthogonal, they can be estimated and re-prepared without introducing any disturbance (or just introducing a vanishing disturbance, in the infinite dimensional case).  This is the reason why the protocols of Refs. \cite{pingpong,mancini,pirandola} necessarily require the random switching between a communication mode and a control mode.  
The present analysis is instead relevant for the analysis of the two-way protocol of Ref. \cite{clonunit}, which uses two mutually unbiased bases of orthogonal qubit  unitaries, given by $\mathcal B_1  =\{\sigma_\mu\}_{\mu =0}^3$ and $\mathcal B_2 =\{  U  \sigma_{\mu}\}_{\mu = 0}^3$, where $\sigma_0 = I$, $\{\sigma_k\}_{k=1}^3$ are the three Pauli matrices, and $U =   ( I +  i  \sum_{k=1}^3   \sigma_k )/2 $ is the rotation of $2\pi/3$  around the axis $\mathbf{n}  = 1/\sqrt 3( 1,1,1)$.  In this case, using the optimal network is a non-trivial eavesdropping attack.  Of course, since the protocol does not involve all possible qubit unitaries,  the optimal trade-off curve for the restricted set  $\mathcal B_1 \cup \mathcal B_2$  could possibly be more favourable to the eavesdropper than the universal trade-off curve derived in this paper.  The analysis of the trade-off for non-universal sets of unitary transformations and the study of the relations between information-disturbance trade-off and quantum cloning for unitary transformations \cite{clonunit} are interesting directions of future research.     

\par {\em Acknowledgments.---} 
We acknowledge an anonymous Referee for useful comments and suggestions.  
This work is supported by
Italian Ministry
of Education through grant PRIN 2008 and
 the EC through  project COQUIT.
Research at
Perimeter Institute for Theoretical Physics is supported
in part by the Government of Canada through NSERC
and by the Province of Ontario through MRI.

\end{document}